# MIT SuperCloud Portal Workspace: Enabling HPC Web Application Deployment

Andrew Prout, William Arcand, David Bestor, Bill Bergeron, Chansup Byun, Vijay Gadepally, Matthew Hubbell,
Michael Houle, Michael Jones, Peter Michaleas, Lauren Milechin, Julie Mullen, Antonio Rosa, Siddharth Samsi,
Albert Reuther, Jeremy Kepner

MIT Lincoln Laboratory, Lexington, MA, U.S.A.

*Abstract*—The MIT SuperCloud Portal Workspace enables the secure exposure of web services running on high performance computing (HPC) systems. The portal allows users to run any web application as an HPC job and access it from their workstation while providing authentication, encryption, and access control at the system level to prevent unintended access. This capability permits users to seamlessly utilize existing and emerging tools that present their user interface as a website on an HPC system creating a portal workspace. Performance measurements indicate that the MIT SuperCloud Portal Workspace incurs marginal overhead when compared to a direct connection of the same service.

## I. INTRODUCTION

The merger of traditional HPC systems, big data clouds, and elastic computing clouds has highlighted new challenges in managing a combined SuperCloud. One key challenge is addressing the need to run web applications and provide the end-users access to them [Kwan 1995, Thomas 2000, Alosio 2001, Milne 2009, Cholia 2010, Atwood 2016, Schuller 2016]. Web-based applications have become a common deployment method for many application development efforts, such as the Jupyter notebook interface to Python, R, Matlab, Octave, and Julia programming languages. Numerous projects expect their users to run private web servers and interact via the web browser. Enabling these same applications on a high performance computing (HPC) system encounters many challenges that do not exist in the local workstation context. The ability for users to start arbitrary web applications that are sufficiently open to the network for that user to connect to them raises obvious security concerns. The design of most HPC systems, in which compute nodes do not have direct external network connectivity, also complicates this goal.

Traditional solutions for this problem, such as SSH port forwarding, are burdensome to teach to users who are not familiar with them. These solutions also can be a security risk in themselves if the tunnel endpoint on the user's workstation is not correctly set up to limit access to just that user. The tunnel endpoints could easily be accessible to other users on the local network or on the same system if it's shared by multiple users.

Even applications that have security built in do not solve these problems. Such applications either must be integrated into the HPC system to use its existing authentication systems or must set up their own parallel authentication system. If integrated into the HPC system, they are no longer a separate user application, as integration would require privileged access to the HPC system. If standing up their own parallel authentication system, that system must be evaluated against the security posture of the system and separately monitored for attacks, weaknesses, reported flaws, and updates that could affect the overall system security. Neither of these options can scale to meet the goal of permitting users to run arbitrary web applications for their own consumption.

We have addressed these challenges by building a Portal Workspace technology capable of dynamically forwarding web applications. The portal acts as a reverse proxy to the internal network of the HPC system that is capable of being dynamically reconfigured by end users to expose their web applications to the external network. It handles encryption, authentication, and authorization of specific users' access to exposed applications.

The organization of the rest of this paper is as follows. Section II describes the technologies used to create the MIT SuperCloud Portal Workspace. Section III describes some examples of efforts enabled by this technology. Section IV shows the performance results and overhead of the system. Section V describes future work in this area. Section VI summarizes the results.

## II. TECHNOLOGIES

The MIT SuperCloud Portal Workspace builds on a number of existing technologies for operating and managing a large, heterogeneous HPC system. These include the software stack, the scheduler, firewall, as well as the Portal Workspace itself.

### A. MIT SuperCloud

The MIT SuperCloud software stack enables traditional enterprise computing and cloud computing workloads to be run on an HPC cluster [Reuther 2013] (see Figure 1). The software stack runs on many different HPC clusters based on a variety of hardware technologies. It supports systems with 10 GigE, and FDR InfiniBand or Intel OmniPath running Internet Protocol over InfiniBand [RFC 4391].

The MIT SuperCloud software stack, which contains all the system and application software, resides on every node. Hosting the application software on each node accelerates the

This material is based upon work supported by the National Science Foundation under Grant No. DMS-1312831. Any opinions, findings, and conclusions or recommendations expressed in this material are those of the author(s) and do not necessarily reflect the views of the National Science Foundation.

launch of large applications (such as databases) and minimizes their dependency and the load on the central storage.

*B. SLURM Resource Manager*

HPC systems require efficient mechanisms for rapidly identifying available computing resources, allocating those resources to programs, and launching the programs on the allocated resources. The open-source SLURM software (https://slurm.schedmd.com) provides these services and is independent of programming language (C, Fortran, Java, Matlab, etc.) or parallel programming model (message passing, distributed arrays, threads, map/reduce, etc.).

Given a range of ports specified in the SLURM configuration, SLURM can assign a requested number of them to jobs. By using the port number assigned by SLURM, users who launch network-enabled jobs, such as web applications, can ensure that no two jobs running on the same compute node attempt to use the same local port number.

*C. User-Based Firewall*

HPC systems traditionally allow their users unrestricted use of their internal network. While this network is normally controlled enough to guarantee privacy without the need for encryption, it does not provide a method to authenticate peer connections. Applications running on this internal network, such as web applications, must provide their own authentication. However, this requirement assumes the application developer has included support and the user will enable these services. This is often not done in practice, especially when an externally developed application does not include the option or the application is an early prototype.

The User-Based Firewall for HPC systems [Prout 2016] solves these issues by enforcing user authentication for all applications at the system level. Using callouts from the Linux netfilter firewall, the process owner of the connecting and listening processes is compared, and the connection is only allowed if the user matches or the connector is a member of the primary group of the listener. On a system that implements the user private group scheme, this makes all network connections private to only the user by default.

*D. MIT SuperCloud Workspace Portal*

The MIT SuperCloud Workspace Portal technology grew out of the prototype built for Defense Research and Engineering (DR&E) [Reuther 2010]. Some of the key requirements for this prototype were the ability to authenticate with a Department of Defense (DoD) standard-issue PKI-enabled smartcard known as the Common Access Card (CAC), to mount the central filesystem via the WebDav protocol [RFC 4918], and to launch HPC jobs without using network ports that are likely to be restricted by firewalls.

The MIT SuperCloud Workplace Portal is based on Apache httpd. It uses a custom Multi-Processing Module (MPM) to impersonate the authenticated user for all operating system calls. This impersonation was necessary for filesystem access through the mod_dav module, which provides WebDav services, to properly respect normal filesystem permissions. This impersonation also enables all other aspects of Apache, including normal web content serving, server-side script execution such as cgi or php scripts, and the watcher module described in our 2010 publication, to act as the authenticated user. This impersonation also ensures that users' actions through the portal are limited to nothing more than they could perform if they opened a SSH session to the system.

We have continued to maintain these capabilities and enhanced the system with the ability to arbitrarily forward to web applications running within the HPC cluster using mod_proxy_http for normal web traffic and mod_proxy_ws for WebSocket traffic [RFC 6455]. By using mod_rewrite's ability to query external applications via the RewriteMap directive, we are able to dynamically look up the intended destination for requests with a prefix that triggers the forwarding system. These two rewrite map helpers, described in more detail below, are also able to perform additional checks prior to allowing the connection to proceed. We have integrated both helpers with the User-Based Firewall, as they run as a system service and would otherwise be exempt, to ensure that users can only access services they started or are authorized to connect to. Along with the ability to launch jobs through the portal, these capabilities allow users to start and immediately connect to web services running on the HPC system.

The first rewrite map helper is used with the address pattern */fw/forward-name/*. The rewrite map helper for this prefix looks for a file in a specific directory with the forward-name. This directory's configuration is similar to the system temporary directory so that any user may create a file, but only the owner of a file may delete it. This setup in effect creates a "first to use it" reservation system for names; as long as the user does not delete the file, the name is reserved for that user. The content of the file is the destination of the forward, but may be truncated to an empty file to disable the forward while maintaining the name reservation. The ability to access the forward is controlled by the *execute* permission on this file, allowing for management through all the normal operating system access control methods.

The file-based forwarding system was originally created prior to the user-based firewall work and was integrated with the virtual smartcard [Prout 2012] to provide a mechanism for protecting against users setting up forwards to unrelated processes. However, the user was required to set up TLS on their backend web application and check for the client certificate. As stated in our 2016 user-based firewall paper, we found that this was not a practical expectation. This forwarding system has now been integrated with the user-based firewall, and the target process must be owned by the same user as the forward file, or group-owned by the same group, to prevent this type of cross-connection without the need for special configuration of the backend web application.

The second rewrite map helper is used with the address pattern */fw2/node:port/*. The rewrite map helper for this prefix simply forwards to the node and port specified. However, before allowing the connection, it enforces the rules articulated for the user-based firewall: that the connecting user must own the web application process, or must be a member of the primary group of the web application process. This simpler method is ideal for applications that do not need a long-lived name or complex access permissions.

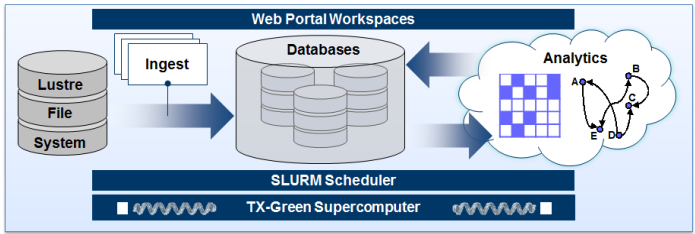

Figure 1. The MIT SuperCloud database system architecture consists of seven components. (1) Lustre parallel file system for high performance file I/O, (2) D4M & pMatlab ingest processes, (3) Accumulo or SciDB parallel database, (4) D4M & pMatlab analytic processes, (5) Accumulo web monitor page viewable through the web portal workspaces, (6) SLURM scheduler for allocating processes to hardware, and (7) the TX-Green

### III. ENABLED APPLICATIONS

This section describes a number of applications that have been enabled by the MIT SuperCloud Workspace Portal. These applications span a diverse range of technologies that includes big data databases, virtual machines, and multi-lingual interactive development environments.

#### A. MIT SuperCloud Database Management System

The MIT SuperCloud database management system [Prout 2015] allows for rapid creation and flexible execution of a variety of the latest scientific databases, including Apache Accumulo and SciDB. It is designed to permit these databases to run on an HPC platform as seamlessly as any other HPC job. The graphical user interface of the database management system is served out through the portal.

The database management system status web page uses the portal's ability to impersonate the connected user to run a php script to generate the status page. Since the script is executed as the connected user, no data can be read or action taken by this script that the user could not access or perform already if logged into the system. No separate logic to implement any filtering for security reasons was necessary, greatly simplifying both the code and the review and approval process for production deployment.

This script generates a list of available databases based on listing the directory containing the databases and reading the "dbinfo" file in each. A database is listed if the dbinfo file is readable to the user according to normal file system permission logic. Database actions, such as start, stop or checkpoint, are submitted to the php script and translated into the shell commands db_start, db_stop, or db_checkpoint, and eventually into calls to the SLURM HPC scheduler as the authenticated user.

The status of each database is read by client-side javascript, so the status can be refreshed without refreshing the whole page. This refresh is done by using asynchronous calls to read the contents of the database's "status" file through the WebDav interface of the portal that exposes the entire HPC filesystem.

The database status page is exposed through a special prefix and rewrite mapper used with the address pattern /dbfw/*database-name*/. The rewrite map helper for this prefix ensures that the user has access to the specified database directory before forwarding to the appropriate compute node and port number for the database's running location and type. The returned HTML is also rewritten to ensure links on the returned page include the needed prefix to be forwarded.

Using this system, researchers have set a number of world records in database performance [Kepner 2014, Samsi 2016]. The database management system played a critical role in setting these records by allowing the researchers to rapidly configure, start, and stop many databases at different scales.

#### B. MIT SuperCloud Virtal Machine System

The MIT SuperCloud Virtual Machine (VM) system [Reuther 2012] allows for the creation of virtual machines that are launched as jobs on the HPC cluster. Running virtual machines on an HPC system is appealing for the ability to rapidly scale out system testing or prototyping, or for the support of legacy operating systems.

These VMs are run using qemu and Linux Kernel-based Virtual Machine (https://www.linux-kvm.org) on top of the MIT SuperCloud system image and can share a compute node side by side with other non-VM jobs. These VMs can utilize Virtual Distributed Ethernet [Davoli 2005] and VXLAN [RFC 7348] to create private networks between VMs and VirtFS folder sharing based on Plan 9 [Jujjuri 2010] to mount the HPC storage through the hypervisor as the user that launched the VM. With these building blocks, a user can simulate an entire enterprise of interconnected desktops, servers and networks, or run a computational program that requires a legacy operating system with minimal sacrifice of access to the large HPC storage.

The graphical user interface of the VM system is served out through the portal. Users can create or edit "plan" files that define their VMs, including the amount and distribution of HPC resources to request, the virtual hardware for each VM, and connections to host folders or virtual networks. These plan files are read and saved through the WebDav interface of the portal that exposes the entire HPC filesystem. Starting and stopping VMs can also be triggered through this web interface to a php script and translated into the shell commands vm_start or vm_stop, and eventually into calls to the HPC scheduler as the authenticated user.

Access to the virtual console of the VMs is provided through the SPICE project (https://www.spice-space.org) and enabled through the portal via the available html5 viewer. A link is displayed for each running VM to launch the viewer and automatically connect to the appropriate compute node and port over WebSockets via the /fw2/node:port/ forwarding prefix. The User-Based Firewall integration of this forwarding prefix prevents unauthorized users from connecting.

#### C. Jupyter Notebook

Jupyter Notebook is a web-based framework for authoring executable code in many languages, optionally with documentation attached alongside the executable code. The executable code can be executed by various *notebook kernels* loaded by the framework. There are over 80 different notebook kernels available on the community supported list (https://github.com/jupyter/jupyter/wiki/Jupyter-kernels). In addition to providing a graphical user interface for some languages that have not previously had one, it also enables a web-based client alternative for others, such as Octave, R, and Matlab (see Figure 2).

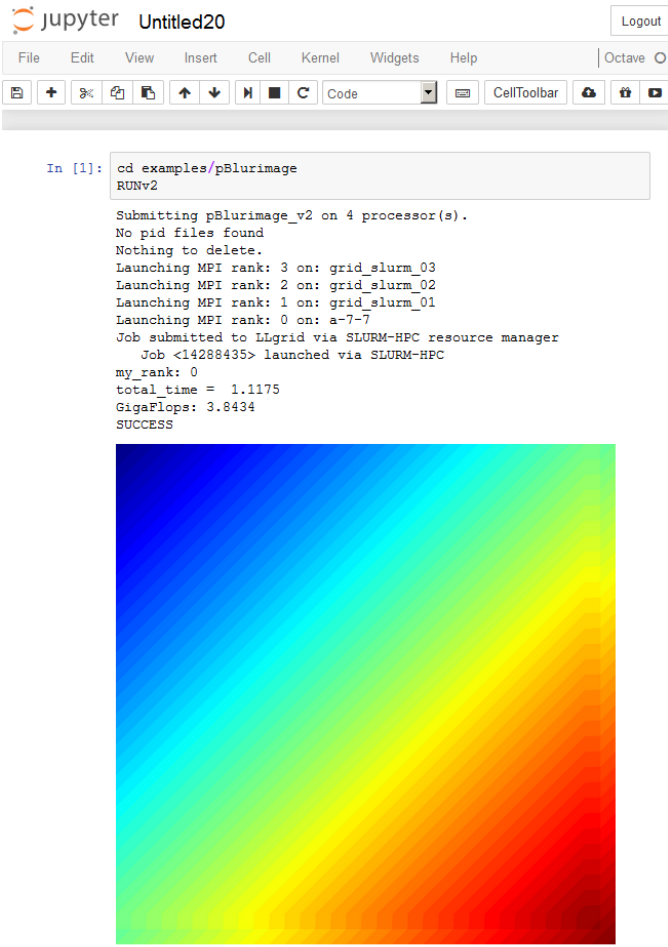

Figure 2. Octave running the pBlurimage example from the pMatlab package from Jupyter Notebook on four HPC nodes.

The standard way to deploy multi-user Jupyter Notebook access is to use the JupyterHub project. However, given our existing work with the MIT SuperCloud Portal, we chose to integrate launching Jupyter Notebook servers into our existing capability. Recent versions of Jupyter Notebook have token authentication turned on by default, which upon startup generates a random string that must be presented to authenticate to the Notebook. Even though the User-Based Firewall rules enforced by the forwarding system would have been sufficient to prevent any user other than the owner of that Notebook from connecting, we chose not to disable this feature. This choice was primarily out of concern that running a Notebook with no apparent authentication may be depreciated in a future release.

The status page of our Jupyter Notebook capability allows a user to start a Jupyter Notebook as an HPC job and display any existing Notebook jobs running as that user. The php script will ssh to the compute node running the Jupyter Notebook job using the user's ssh key to query the token for that Notebook and display an appropriate link to connect that includes the /fw2/node:port/ prefix and the token suffix. The launch action is submitted to the php script and translated into an HPC scheduler command to dispatch a job to start a Jupyter Notebook instance as the authenticated user.

By choosing to use our portal capability, we avoided having another service with user impersonation rights. Vetting the JupyterHub codebase and configuration to ensure it interacted correctly on our system would not have been a small undertaking. Additionally, as with any service running with elevated privileges, we would have had to monitor the project for vulnerability and patch announcements that would have affected the security of our systems. Using our own implementation, we also have more freedom to expose more complex SLURM options for GPU and Intel Xeon Phi Knights Landing compute nodes than easily allowed by the JupyterHub SLURM spawner's static configurations.

IV. PERFORMANCE

The performance of the portal can be measured by comparing how fast a web application is when accessed directly vs. through the portal. Previous work on thin clients, which modern web applications often are, informs us that latency is the critical performance metric for evaluating the user experience [Tolia 2006].

We have benchmarked two measures of latency: overall page load time and the load time of each individual http request. The test was performed for loading the initial user interface of Jupyter Notebook, which results in 31 separate http requests being performed. We performed these tests on an HPC system utilizing 10GigE for its interconnection and with a

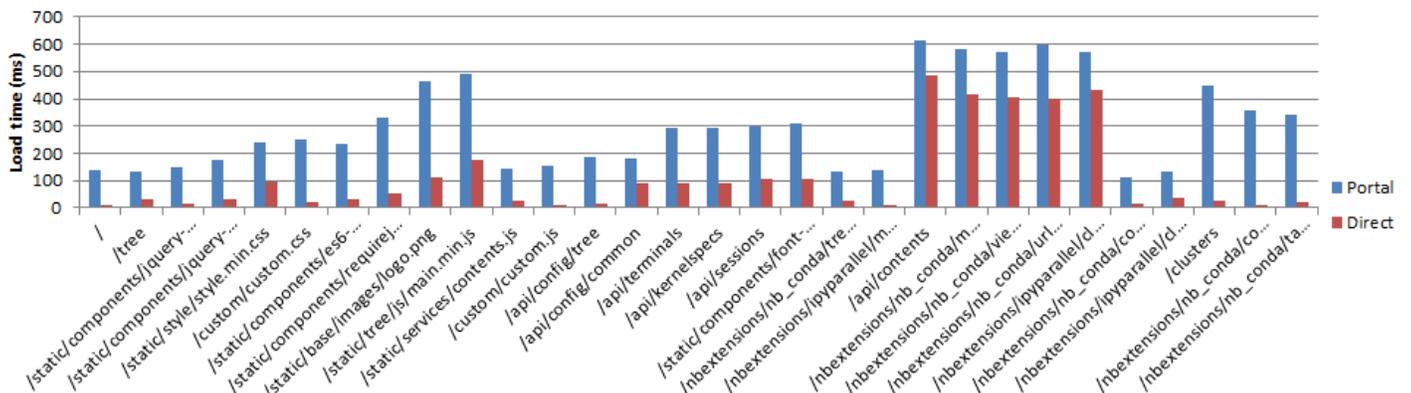

Figure 3. Comparison of the load times of the 31 http requests needed to load Jupyter Notebook compared for direct connection vs. proxied via the portal.

10GigE link to the workstation used to perform the tests. To measure the page load times, Firefox's Network Monitor was used.

The overall page load time was 3,430ms when connecting directly, and 4,494ms when connecting through the portal. However, each of the 31 http requests showed a different amount of delay, shown in Figure 3. The delays range from 87ms to 420ms depending on the size and type of the content. The type of content is relevant to the need for the proxy to look for URLs in the content that may need to be rewritten to the proxy's URL space instead of the backend server's URL space.

The sum of the delays encountered by each of the 31 http requests, 5,691ms, is significantly greater than the actual delay of the loading process. This difference is to be expected as many of the http requests are performed in parallel. However, the parallelization of the requests would have been hampered by the serial nature of the RewriteMap inter-process communication.

Feedback from users has indicated that this latency is quite acceptable and is well worth it to them in order to obtain a zero-admin web interface to an HPC system.

V. FUTURE WORK

In the future, we will concentrate on enabling more applications to be exposed through the portal and building better user tools to automate starting common web applications as portal-enabled HPC jobs.

Future work will also look into the root causes of the latency of the portal and seek to reduce them. Implementing a short-lived cache for the ident2 component of the user-based firewall should significantly reduce the delays for applications that issue multiple requests in quick succession, as seen in the example we examined in Section IV. We will also look into integrating the RewriteMap helpers as native Apache httpd modules to eliminate the inter-process communication serialization as a source of the overhead they could now be experiencing.

VI. SUMMARY

The MIT SuperCloud Portal allows users to seamlessly access web-based applications running as jobs on an HPC system. It has enabled users to run applications that present their user interface as a web site without the complications, and potential issues, of using port forwarding. We have made available options for users to connect to applications they launch by either a registered name or by knowing the node and port number that their job was assigned. With the addition of Jupyter Notebook, our portal technology is becoming a fully featured workspace for high performance computing.

REFERENCES

[Alosio 2001] Alosio, G., Cafaro, M., Kesselman, C. and Williams, R., 2001. "Web access to supercomputing". *Computing in Science & Engineering*, *3*(6), pp.66-72.

[Atwood 2016] C. Atwood, R. Goebbert, J. Calahan, T. Hromadka, T. Proue, W. Monceaux, and J. Hirata, J., 2016. "Secure Web-Based Access for Productive Supercomputing," *Computing in Science & Engineering*, *18*(1), pp.63-72.

[Cholia 2010] S. Cholia, D. Skinner, and J. Boverhof, 2010, November. "NEWT: A RESTful service for building High Performance Computing web applications." In *Gateway Computing Environments Workshop (GCE), 2010* (pp. 1-11). IEEE.

[Davoli 2005] R. Davoli, "VDE: Virtual Distributed Ethernet," *First International Conference on Testbeds and Research Infrastructures for the Development of Networks and Communities (Tridentcom 2005)*, Trento, Italy, Feb 2005.

[Jujjuri 2010] V. Jujjuri, E. Van Hensbergen, A. Liguori, B. Pulavarty, "Virtfs–a virtualization aware file system pass-through," *Ottawa Linux Symposium (OLS)*, pp. 109-120, July 2010.

[Kepner 2014] J. Kepner, W. Arcand, D. Bestor, B. Bergeron, C. Byun, V. Gadepally, M. Hubbell, P. Michaleas, J. Mullen, A. Prout, A. Reuther, A. Rosa, and C. Yee, 2014, September. "Achieving 100,000,000 database inserts per second using Accumulo and D4M." In *High Performance Extreme Computing Conference (HPEC), 2014 IEEE* (pp. 1-6). IEEE.

[Kwan 1995] T. Kwan, R. McGrath, and D. Reed, 1995. "NCSA's world wide web server: Design and performance." *Computer*, *28*(11), pp.68-74.

[Milne 2009] I. Milne, D. Lindner, M. Bayer, D. Husmeier, G. McGuire, D. Marshall, and F. Wright, 2009. "TOPALi v2: a rich graphical interface for evolutionary analyses of multiple alignments on HPC clusters and multi-core desktops." *Bioinformatics*, *25*(1), pp.126-127.

[Prout 2012] A. Prout, W. Arcand, D. Bestor, B. Bergeron, M. Hubbell, J. Kepner, A. McCabe, P. Michaleas, J. Mullen, A. Reuther, and A. Rosa, "Scalable Cryptographic Authentication for High Performance Computing," *IEEE High Performance Extreme Computing (HPEC) Conference*, Sep 10-12, 2012, Waltham, MA.

[Prout 2015] A. Prout, J. Kepner, P. Michaleas, W. Arcand, D. Bestor, D. Bergeron, C. Byun, L. Edwards, V. Gadepally, M. Hubbell, J. Mullen, A. Rosa, C. Yee, and A. Reuther, "Enabling On-Demand Database Computing with MIT SuperCloud Database Management System," *IEEE High Performance Extreme Computing (HPEC) Conference,* Sep 15-17, 2015, Waltham, MA.

[Prout 2016] A. Prout, W. Arcand, D. Bestor, D. Bergeron, C. Byun, V. Gadepally, M. Hubbell, M. Houle, M. Jones, P. Michaleas, L. Milechin, J. Mullen, A. Rosa, S. Samsi, A. Reuther, and J. Kepner, "Enhancing HPC Security with a User-Based Firewall," *IEEE High Performance Extreme Computing (HPEC) Conference*, Sep 13-15, 2016, Waltham, MA.

[Reuther 2010] A. Reuther, W. Arcand, C. Byun, B. Bergeron, M. Hubbell, J. Kepner, A. McCabe, P. Michaleas, J. Mullen, and A. Prout, "DR&E LLGrid Portal: Interactive Supercomputing for DoD," *2010 High Performance Embedded Computing Workshop*, Sep 15-16, 2010, Lexington, MA.

[Reuther 2012] A. Reuther, P. Michaleas, A. Prout, and J. Kepner, "HPC-VMs: Virtual Machines in High Performance Computing Systems," *IEEE High Performance Extreme Computing (HPEC) Conference*, Sep 10-12, 2012, Waltham, MA.

[Reuther 2013] A. Reuther, J. Kepner, W. Arcand, D. Bestor, B. Bergeron, C. Byun, M. Hubbell, P. Michaleas, J. Mullen, A. Prout, and A. Rosa, "LLSuperCloud: Sharing HPC Systems for Diverse Rapid Prototyping," *IEEE High Performance Extreme Computing (HPEC) Conference,* Sep 10-12, 2013, Waltham, MA.

[RFC 4391] J. Chu and V. Kashyap, "Transmission of IP over InfiniBand (IPoIB)," RFC 4391, DOI 10.17487/RFC4391, April 2006, <http://www.rfc-editor.org/info/rfc1413>.

[RFC 4918] L. Dusseault, "HTTP Extensions for Web Distributed Authoring and Versioning (WebDAV)," RFC 4918, DOI 10.17487/RFC4918, June 2007, <https://tools.ietf.org/html/rfc4918>.

[RFC 6455] I. Fette and A. Melnikov, "The WebSocket Protocol," RFC 6455, DOI 10.17487/RFC6455, December 2011, <https://tools.ietf.org/html/rfc6455>.

[RFC 7348] M. Mahalingam, D. Dutt, K. Duda, P. Agarwal, L. Kreeger, T. Sridhar, M. Bursell, and C. Wright, "Virtual eXtensible Local Area Network (VXLAN): A Framework for Overlaying Virtualized Layer 2 Networks over Layer 3 Networks," RFC 7348, DOI 10.17487/RFC7348, August 2011, <https://tools.ietf.org/html/rfc7348>.

[Samsi 2016] S. Samsi, L. Brattain, W. Arcand, D. Bestor, B. Bergeron, C. Byun, V. Gadepally, M. Hubbell, M., Jones, A. Klein, A, P. Michaleas, J. Mullen, A. Prout, A. Rosa, C. Yee, and A. Reuther. "Benchmarking SciDB data import on HPC systems." In *High Performance Extreme Computing Conference (HPEC), 2016 IEEE* (pp. 1-5). IEEE.

[Schuller 2016] B. Schuller, K. Benedyczak, R. Grunzke, M. Petrova-ElSayed, and J. Rybicki, J., 2016. "UNICORE 7-Middleware Services for Distributed and Federated Computing." In *International Conference on High Performance Computing & Simulation (HPCS)* (No. FZJ-2016-04344). Jülich Supercomputing Center.

[Thomas 2000] M. Thomas, S. Mock, and J. Boisseau, 2000. "Development of Web toolkits for computational science portals: The NPACI HotPage." In *High-Performance Distributed Computing, 2000. Proceedings. The Ninth International Symposium on* (pp. 308-309). IEEE.

[Tolia 2006] N. Tolia, D. G. Andersen, and M. Satyanarayanan, "Quantifying Interactive User Experience on Thin Clients," *IEEE Computer*, vol. 39, no. 3, pp. 46-52, Mar. 2006.